\documentclass[
 amsmath,amssymb,
 reprint,
]{revtex4-1}

\usepackage[dvips]{graphicx}
\usepackage{dcolumn}
\usepackage{bm}
\usepackage[]{hyperref}
\hypersetup{
    colorlinks=true, 
    linkcolor=blue,
    citecolor=blue,
    filecolor=blue,
    urlcolor=blue,
    }
\usepackage[all]{hypcap}

\begin{document}

\title[]{Fast pick up technique for high quality heterostructures of bilayer graphene and hexagonal boron nitride}

\author{P. J. Zomer}
\email{pj.zomer@gmail.com}
\affiliation{Physics of Nanodevices, Zernike Institute for Advanced Materials, University of Groningen, Groningen, The Netherlands
}

\author{M. H. D. Guimar\~{a}es}
\affiliation{Physics of Nanodevices, Zernike Institute for Advanced Materials, University of Groningen, Groningen, The Netherlands
}

\author{J. C. Brant}
\affiliation{Physics of Nanodevices, Zernike Institute for Advanced Materials, University of Groningen, Groningen, The Netherlands
}

\author{N. Tombros}
\affiliation{Physics of Nanodevices, Zernike Institute for Advanced Materials, University of Groningen, Groningen, The Netherlands
}

\author{B. J. van Wees}
\affiliation{Physics of Nanodevices, Zernike Institute for Advanced Materials, University of Groningen, Groningen, The Netherlands
}
\date{\today}

\begin{abstract}

We present a fast method to fabricate high quality heterostructure devices by picking up crystals of arbitrary sizes. 
Bilayer graphene is encapsulated with hexagonal boron nitride to demonstrate this approach, showing good electronic quality with mobilities ranging from 17~000~cm$^{2}$V$^{-1}$s$^{-1}$ at room temperature to 49~000~cm$^{2}$V$^{-1}$s$^{-1}$ at 4.2~K, and entering the quantum Hall regime below 0.5~T.
This method provides a strong and useful tool for the fabrication of future high quality layered crystal devices.

\end{abstract}

\maketitle

A critical step for high mobility graphene device fabrication and the rising field of van der Waals heterostructures\cite{GeimVdWHeteroStruc} is marked by the development of polymer based dry transfer methods for two-dimensional (2D) crystals.\cite{TJoPCC112_Reina2008_TransferringandIdentificationofSingle-andFew-LayerGrapheneonArbitrarySubstrates,DeanBNtransfer,Mayorov_um_scale_balltr_GBN,ZomerTransfer} 
With these methods, high quality graphene devices on hexagonal boron nitride (h-BN) and more complicated stacks have become generally accessible, but the early methods face a major setback. 
The method of stacking the crystals one by one typically leaves each transferred crystal contaminated by polymer. 
To obtain a high quality device, thorough cleaning is required before proceeding with device fabrication or measurement. 
This cleaning step typically involves several hours of annealing\cite{DeanBNtransfer,Mayorov_um_scale_balltr_GBN,ZomerTransfer} or it may go as far as nanobrooming the entire graphene flake using contact mode atomic force microscopy (AFM).\cite{JalilianAFMsgs,GoossensAFMclean} 
Altogether this makes the fabrication of a multilayer heterostructure not only very time consuming, but also risky as each step may again introduce contaminants to the stack.

\begin{figure}[b]
\includegraphics[width=85mm]{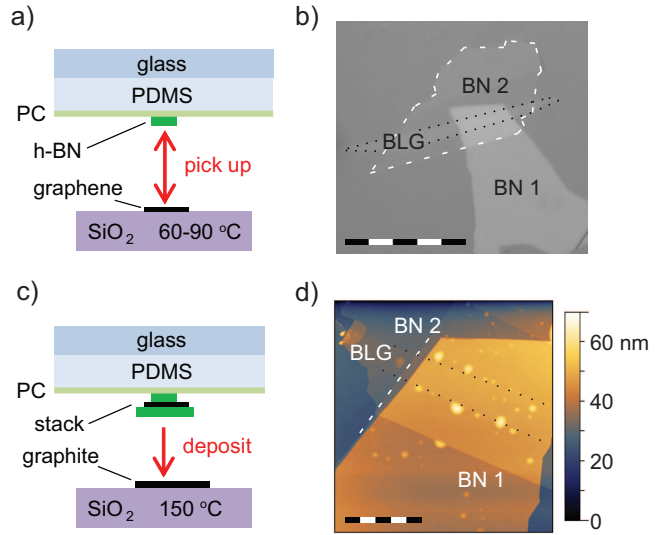}
\caption{
\label{fig:Fig1} (Color online) 
a) Cross section (not to scale) of the glass slide used to pick up a graphene flake. 
The pick up process can be repeated several times to create a multilayer heterostructure.
b) A stack in the making on PC imaged before deposition on a substrate. BN1 was used to pick up the BLG (black short dashed line) and then BN2 (white long dashed line). The scale bar equals 20~$\mu$m.
c) In the last step the stack is released onto graphite on the substrate at an elevated temperature ($\sim$150~$^{\circ}$C).
d) AFM micrograph of the stack in b), deposited on a few layer graphene flake on SiO$_2$. 
Two black dotted lines trace the BLG edges, the white dashed line marks the left edge of the few layer graphene flake.
The scale bar equals 5~$\mu$m.
}
\end{figure} 

This issue has recently been overcome by L. Wang et al.\cite{Wang1Dcontactto2Dmaterial}, introducing a method that allows for polymer free assembly of layered materials based on van der Waals force.
Instead of depositing a 2D crystal, e.g. h-BN, directly on top of another crystal, e.g. graphene, one can use the h-BN to pick up the graphene from the substrate.
This can be done because the van der Waals force between the atomically flat h-BN and graphene is stronger than between the graphene or the h-BN and the rough SiO${_2}$ substrate.
The power of this method lies in the fact that now the interface between the two crystals has not been contaminated by polymer, and one can directly pick up the next crystal.
This way the materials inside the stack not only remain much cleaner, but a stack can also be fabricated considerably faster. 
One problem when using this method is the reduced capability to pick up graphene flakes larger than the used top h-BN crystal.
Therefore one has to etch through the stack before making one-dimensional (1D) contacts to the graphene.
While resulting in very high quality devices with electron mean free paths up to 21~${\mu}$m and good electrical contact,\cite{Wang1Dcontactto2Dmaterial} this limitation can be problematic for certain device types for which 1D contacts are not desirable, e.g. spintronic devices that include tunnel barriers at the contact interface.\cite{ZomerSpinBN}

In this letter we present a method which allows for fabrication of high quality graphene devices encapsulated in h-BN by successively picking up crystals.
The advantage of our approach is that stacks are made without size constraints on the graphene.
No cleaning steps are used in the fabrication process, which considerably increases the fabrication speed of a stack from 1 or 2 days down to half an hour.
The quality we achieve for h-BN encapsulated BLG devices becomes similar to that of current annealed suspended bilayer graphene devices,\cite{VelascoInsBLG,AllenQDinsuspBLG,vanElferen_supsBLG_QHferromagnetism,VeliguraTrGapinBLG0B} while keeping the benefits of having a substrate, for example when including top- and backgates.\cite{GoossensQdot}
In our device, we observe quantum Hall levels to start develop below 0.5~T at 4.2~K and the degeneracy of the first Landau level is broken already around 3~T.
With this newly developed method, high quality graphene devices are readily accessible without the need for a furnace or etching system.

The fabrication of a stack starts with the preparation of a glass slide that is used to pick up other crystals. 
We first prepare a thin film of polycarbonate (PC, Sigma Aldrich, 6\% dissolved in chloroform mixed at HQ Graphene). 
Using a pipette, the PC is dripped on a microscope slide. 
Because the PC is difficult to spincoat, a second glass slide is directly put on top of the PC covered slide, spreading out the PC, after which they are immediately separated by sliding the two slides over each other.
This results in a reasonably uniform PC film which is then left to harden in air for about 15 minutes.

On top of the dried PC film we exfoliate commercially available h-BN (HQ Graphene) by mechanical cleavage using adhesive tape.
By optical microscope we select an h-BN crystal suitable to serve as the top layer for the stack.
Then we take a new glass slide on which we put a $\sim$4$\times$4$\times$1~mm piece of polydimethylsiloxane (PDMS).
The PC with the top layer h-BN is removed from its glass slide using adhesive tape and laid across the PDMS, with the h-BN flake facing up.
Now the newly made slide can be used to pick up other crystals.
A schematic cross section at this stage of the glass slide, PDMS, PC film and h-BN can be seen in the top half of Fig.\ref{fig:Fig1}a.

The crystals that are to be picked up in the next step are prepared by exfoliation of graphite (HOPG) or h-BN on separate Si/SiO$_2$ (500~nm) wafers.
Before exfoliation the Si/SiO$_2$ substrates are cleaned in a furnace at 400~$^{\circ}$C for 5 minutes in air.
By optical contrast we select BLG and thin h-BN (5-20~nm) flakes for our stack.
The glass slide is mounted in an optical mask aligner with the top layer h-BN flake facing down.
The SiO$_2$, containing BLG, is fixed on the chuck of the mask aligner, this way we can accurately align the h-BN to the BLG (Fig.\ref{fig:Fig1}a) and start bringing the PC and the SiO$_2$ in contact.
While closely following the progress by an optical microscope, we continue to make contact until the two target crystals are almost in touch, which can be easily distinguished optically.
Now we heat the chuck to between 60 and 90~$^{\circ}$C.
As the SiO$_2$ heats up, the contact area with the PC consequently gradually increases further.
When contact is made between the h-BN and BLG, we switch the heater off, which in turn causes the PC film to slowly retract from the substrate as it cools down.

When fully retracted, the PC film remains intact on the PDMS and the BLG is picked up.
The process just described can easily be repeated to pick up another crystal, in our case h-BN, to form a multilayer heterostructure.
An optical micrograph of a stack in the making on the PC layer is shown in Fig.\ref{fig:Fig1}b, where we started with BN1 and then subsequently picked up the BLG and BN2. 
Note that due to the good adhesion between graphene and PC, BN1 does not have to entirely cover the BLG for the pick up to be successful.
To release this stack onto a substrate containing graphite that we aim to use as a back gate, we follow the same procedure as for a pick up until the retraction step (Fig.\ref{fig:Fig1}c). 
While still in contact we heat the substrate up to $\sim$150~$^{\circ}$C in order to melt the PC onto it.
Next, the hot chuck is carefully retracted, releasing the PC from the PDMS and leaving us with the desired stack on the SiO$_2$ substrate.
The PC is removed by rinsing in chloroform for 10 minutes.
Overall the pick up process has worked for almost 100\% of the attempts, excluding manual misalignment.
Typically, only a small amount of bubbles is observed in the final stack.\cite{Haigh_NatMat_bubblesGhBN}
Most form just outside the encapsulated graphene area, at the BLG edges, as the AFM image in Fig.\ref{fig:Fig1}d shows.

\begin{figure}[]
\includegraphics[width=85mm]{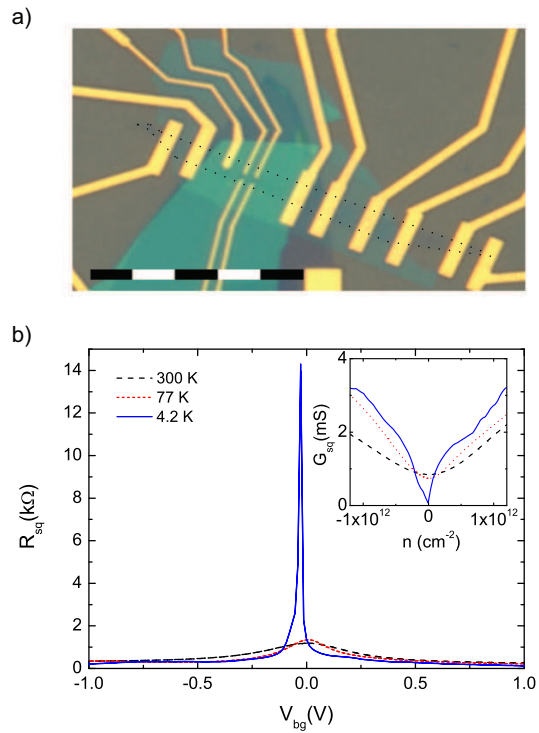}
\caption{
\label{fig:Fig2} (Color online) 
a) A finalized device using the stack imaged in Fig.\ref{fig:Fig1}b and d. 
The black dotted line traces the BLG, the scale bar equals 20~$\mu$m. 
b) Encapsulated bilayer graphene square resistance $R_{sq}$ for different temperatures as a function of the backgate voltage. The inset shows the respective conductance $G_{sq}$ as function of charge carrier density $n$.
}
\end{figure} 

The stack is fabricated into an electronic device using standard electron beam lithography and electron beam evaporation techniques.
A finished device is shown in Fig.\ref{fig:Fig2}a, the contacts are made using of Ti/Au (5/45~nm).
From the substrate up, this stack consists of a few layer graphene flake ($\sim$1.5~nm) to be used as backgate, h-BN ($\sim$5.5~nm), BLG and again h-BN ($\sim$16~nm).
Since we are interested in the quality of the encapsulated region, we also do not require an additional cleaning step after processing the device.
For non-encapsulated graphene such a step would be needed to remove e-beam resist residues which otherwise degrade the electronic quality of the device.

The first characterization of the encapsulated BLG is done by measuring the square resistance $R_{sq}$ as function of backgate voltage $V_{bg}$.
The results, measured at room temperature, 77~K and 4.2~K, are shown in Fig.\ref{fig:Fig2}b.
What can be noted directly, is that the Dirac peak is sharp and very close to zero backgate voltage.
This indicates that the device has little intrinsic doping ($\sim1.5\times10^{11}$~cm$^{-2}$) and low inhomogeneity, as expected for graphene on h-BN.\cite{XueSTM_GBN,Decker_STM_GBN} 
Furthermore, the device has proven to be robust.
After the measurements were taken at room temperature and 77~K the device was removed from the measurement setup, taken off its chip carrier and rewired, before loading it into a cryostat. 
This may have caused the small shift in the Dirac point from 0.01~V at 77~K to -0.025~V at 4.2~K, but otherwise this did not notably degrade the device quality.

\begin{figure}[]
\includegraphics[width=85mm]{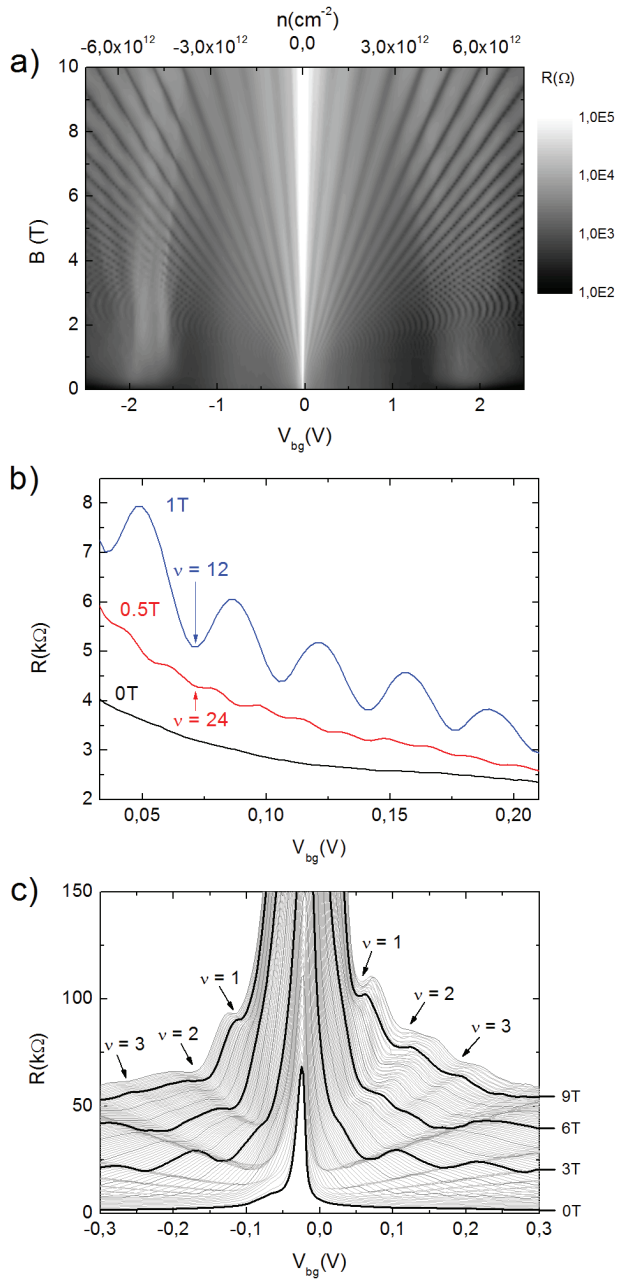}
\caption{
\label{fig:Fig3} (Color online) 
a) Resistance as function of backgate voltage and magnetic field perpendicular to the BLG plane at 4.2~K. At negative and positive gate voltages additional resistance peaks can be distinguished.
b) Traces taken from the data in a) show quantum Hall levels already forming below 0.5~T.
c) Close up of the data in a) around the charge neutrality point. 
The magnetic field for subsequent traces increases in steps of 0.1~T and they are offset with respect to each other by 500~$\Omega$ for clarity. 
The measurements reveal the development of the $\nu=2$ plateau already at 3~T.
}
\end{figure}

From the backgate dependence of the square resistance we can extract the mobility.
First the charge carrier density is calculated using $n=(V-V{_D})\epsilon_{0} \epsilon_{r}/e d$. 
Here $V$ is the applied backgate voltage, $V_{D}$ the charge neutrality voltage (Dirac point), $\epsilon_{0}$ is the permittivity of free space, $\epsilon_{r}$ is the relative permittivity of h-BN, $e$ is the electron charge and $d$ is the h-BN thickness.
When subject to a magnetic field perpendicular to the BLG plane, the device exhibits clear quantum Hall levels, shown in Fig.\ref{fig:Fig3}a.
This provides an alternative way to determine $n$ via the filling factor, expressed as $\nu=nh/eB$, where h is Planck's constant and $B$ is the magnetic field.
Combining both methods we can determine the ratio $\epsilon_{r}/d$ for our device, which we find to be $0.52\pm0.01$~nm$^{-1}$ (using the data from Fig.\ref{fig:Fig3}a).
AFM yields an h-BN thickness $d=5.5\pm0.2$~nm, hence $\epsilon_{r}=2.9\pm0.2$.
Using $\mu=1/neR_{sq}$ we find a room temperature mobility of 17~000~cm$^{2}$V$^{-1}$s$^{-1}$ at $n\approx3.8\times10^{11}$cm$^{-2}$.
For 77~K and 4.2~K we find respectively 30~000~cm$^{2}$V$^{-1}$s$^{-1}$ at $n\approx1.9\times10^{11}$~cm$^{-2}$ and 49~000~cm$^{2}$V$^{-1}$s$^{-1}$ at $n\approx1.4\times10^{10}$~cm$^{-2}$.
The mobility is for all cases determined at the inflection point, where d$R_{sq}$/d$n$ has an extreme.

The quantum Hall data in Fig.\ref{fig:Fig3}a shows, besides a  Landau fan around 0~V, faint additional fans originating at $\pm$1.8~V.
These side peaks are likely a consequence of the Moir\'e superlattice that exists for single and bilayer graphene on h-BN~\cite{YankowitzGBNsuperlatticeSTM,PonomarenkoHBF,HuntHBF,DeanHBF}, having both the same crystallographic lattice with a mismatch of 1.8\%.\cite{PeaseXrayBN} 
The wavelength $\lambda$ of the Moir\'e pattern can be estimated from $n\approx5.2\times10^{12}$~cm$^{-2}$ at which the side peaks occur.
At this point the Moir\'e minibands have become fully occupied, which requires 4 electrons per unit cell for valley and spin degeneracy.
So using $n=4n_{0}$ with $1/n_{0}=\sqrt{3}\lambda^{2}/2$ as the superlattice unit cell area, we obtain $\lambda\sim9.4$~nm.
This corresponds to an angle of $\sim$1.1$^{\circ}$ between the graphene and boron nitride lattices.\cite{YankowitzGBNsuperlatticeSTM} 

The magnetic field data can also be used as an indication of the device quality.
Using $\mu B \approx 1$ as a condition for the device to enter the quantum Hall regime, a mobility can be determined.\cite{SSC146_Bolotin2008_UltrahighElectronMobilityinSuspendedGraphene,Elias_DiracConeReshaped_suspSLG}
We observe the development of quantum Hall at magnetic fields below 0.5~T in Fig.\ref{fig:Fig3}b, indicating that the mobility is at least 20~000~cm$^{2}$V$^{-1}$s$^{-1}$.
This observation is comparable to what has been achieved for suspended BLG.\cite{VelascoInsBLG,AllenQDinsuspBLG,vanElferen_supsBLG_QHferromagnetism,VeliguraTrGapinBLG0B} 
The filling factors are indicated at $\nu=24$ and $\nu=12$ for 0.5~T and 1~T respectively in Fig.\ref{fig:Fig3}b.
For fields around 1.5~T oscillations are still present at filling factors over $\nu=192$.

Furthermore it can be noted in Fig.\ref{fig:Fig3}c that the degeneracy of the lowest Landau level starts breaking below 3~T, forming plateaus with filling factor $\nu=2$.
At higher magnetic fields the degeneracy is lifted further, forming first a plateau at $\nu=3$ around 5~T and later one at $\nu=1$ just below 8~T.
This can also be taken as a demonstration of a good device quality.\cite{vanElferen_supsBLG_QHferromagnetism}

In conclusion we demonstrated an easy device fabrication method using BLG and h-BN that can be used to build complicated stacks of 2D crystals, opening up many opportunities for material engineering. 
The main advantage is that the crystal interfaces remain very clean without the need for any cleaning steps, which additionally makes building stacks considerably faster.
Since our approach allows for picking up crystals of any size, no etching is required to access any layer in the stack.
We showed measurements on a BLG encapsulated in h-BN, fabricated using this method.
The quality of this device is comparable to suspended BLG devices, as can be seen from regular charge transport and quantum Hall measurements.
With the current need for high quality graphene devices and the trend towards other 2D materials, this method is a strong and important tool for fabrication.

We thank L. Wang, I. Meric, C. R. Dean and P. Kim for helpful discussion and we acknowledge B. Wolfs, J. G. Holstein, H. M. de Roosz and H. Adema for their technical assistance. 
The research leading to these results has received funding from the European Union Seventh Framework Programme under grant agreement n$^{\circ}$604391 Graphene Flagship, the Dutch Foundation for Fundamental Research on Matter (FOM), NWO, NanoNed, the Zernike Institute for Advanced Materials and CNPq, Brazil.


\begin{thebibliography}{24}%
\makeatletter
\providecommand \@ifxundefined [1]{%
 \@ifx{#1\undefined}
}%
\providecommand \@ifnum [1]{%
 \ifnum #1\expandafter \@firstoftwo
 \else \expandafter \@secondoftwo
 \fi
}%
\providecommand \@ifx [1]{%
 \ifx #1\expandafter \@firstoftwo
 \else \expandafter \@secondoftwo
 \fi
}%
\providecommand \natexlab [1]{#1}%
\providecommand \enquote  [1]{``#1''}%
\providecommand \bibnamefont  [1]{#1}%
\providecommand \bibfnamefont [1]{#1}%
\providecommand \citenamefont [1]{#1}%
\providecommand \href@noop [0]{\@secondoftwo}%
\providecommand \href [0]{\begingroup \@sanitize@url \@href}%
\providecommand \@href[1]{\@@startlink{#1}\@@href}%
\providecommand \@@href[1]{\endgroup#1\@@endlink}%
\providecommand \@sanitize@url [0]{\catcode `\\12\catcode `\$12\catcode
  `\&12\catcode `\#12\catcode `\^12\catcode `\_12\catcode `\%12\relax}%
\providecommand \@@startlink[1]{}%
\providecommand \@@endlink[0]{}%
\providecommand \url  [0]{\begingroup\@sanitize@url \@url }%
\providecommand \@url [1]{\endgroup\@href {#1}{\urlprefix }}%
\providecommand \urlprefix  [0]{URL }%
\providecommand \Eprint [0]{\href }%
\providecommand \doibase [0]{http://dx.doi.org/}%
\providecommand \selectlanguage [0]{\@gobble}%
\providecommand \bibinfo  [0]{\@secondoftwo}%
\providecommand \bibfield  [0]{\@secondoftwo}%
\providecommand \translation [1]{[#1]}%
\providecommand \BibitemOpen [0]{}%
\providecommand \bibitemStop [0]{}%
\providecommand \bibitemNoStop [0]{.\EOS\space}%
\providecommand \EOS [0]{\spacefactor3000\relax}%
\providecommand \BibitemShut  [1]{\csname bibitem#1\endcsname}%
\let\auto@bib@innerbib\@empty
\bibitem [{\citenamefont {Geim}\ and\ \citenamefont
  {Grigorieva}(2013)}]{GeimVdWHeteroStruc}%
  \BibitemOpen
  \bibfield  {author} {\bibinfo {author} {\bibfnamefont {A.~K.}\ \bibnamefont
  {Geim}}\ and\ \bibinfo {author} {\bibfnamefont {I.~V.}\ \bibnamefont
  {Grigorieva}},\ }\href {\doibase 10.1038/nature12385} {\bibfield  {journal}
  {\bibinfo  {journal} {Nature}\ }\textbf {\bibinfo {volume} {499}},\ \bibinfo
  {pages} {419} (\bibinfo {year} {2013})}\BibitemShut {NoStop}%
\bibitem [{\citenamefont {Reina}\ \emph {et~al.}(2008)\citenamefont {Reina},
  \citenamefont {Son}, \citenamefont {Jiao}, \citenamefont {Fan}, \citenamefont
  {Dresselhaus}, \citenamefont {Liu},\ and\ \citenamefont
  {Kong}}]{TJoPCC112_Reina2008_TransferringandIdentificationofSingle-andFew-LayerGrapheneonArbitrarySubstrates}%
  \BibitemOpen
  \bibfield  {author} {\bibinfo {author} {\bibfnamefont {A.}~\bibnamefont
  {Reina}}, \bibinfo {author} {\bibfnamefont {H.}~\bibnamefont {Son}}, \bibinfo
  {author} {\bibfnamefont {L.}~\bibnamefont {Jiao}}, \bibinfo {author}
  {\bibfnamefont {B.}~\bibnamefont {Fan}}, \bibinfo {author} {\bibfnamefont
  {M.~S.}\ \bibnamefont {Dresselhaus}}, \bibinfo {author} {\bibfnamefont
  {Z.}~\bibnamefont {Liu}}, \ and\ \bibinfo {author} {\bibfnamefont
  {J.}~\bibnamefont {Kong}},\ }\href {http://dx.doi.org/10.1021/jp807380s}
  {\bibfield  {journal} {\bibinfo  {journal} {The Journal of Physical Chemistry
  C}\ }\textbf {\bibinfo {volume} {112}},\ \bibinfo {pages} {17741} (\bibinfo
  {year} {2008})}\BibitemShut {NoStop}%
\bibitem [{\citenamefont {Dean}\ \emph {et~al.}(2010)\citenamefont {Dean},
  \citenamefont {Young}, \citenamefont {Meric}, \citenamefont {Lee},
  \citenamefont {Wang}, \citenamefont {Sorgenfrei}, \citenamefont {Watanabe},
  \citenamefont {Taniguchi}, \citenamefont {Kim}, \citenamefont {Shepard},\
  and\ \citenamefont {Hone}}]{DeanBNtransfer}%
  \BibitemOpen
  \bibfield  {author} {\bibinfo {author} {\bibfnamefont {C.~R.}\ \bibnamefont
  {Dean}}, \bibinfo {author} {\bibfnamefont {A.~F.}\ \bibnamefont {Young}},
  \bibinfo {author} {\bibfnamefont {I.}~\bibnamefont {Meric}}, \bibinfo
  {author} {\bibfnamefont {C.}~\bibnamefont {Lee}}, \bibinfo {author}
  {\bibfnamefont {L.}~\bibnamefont {Wang}}, \bibinfo {author} {\bibfnamefont
  {S.}~\bibnamefont {Sorgenfrei}}, \bibinfo {author} {\bibfnamefont
  {K.}~\bibnamefont {Watanabe}}, \bibinfo {author} {\bibfnamefont
  {T.}~\bibnamefont {Taniguchi}}, \bibinfo {author} {\bibfnamefont
  {P.}~\bibnamefont {Kim}}, \bibinfo {author} {\bibfnamefont {K.~L.}\
  \bibnamefont {Shepard}}, \ and\ \bibinfo {author} {\bibfnamefont
  {J.}~\bibnamefont {Hone}},\ }\href {\doibase 10.1038/nnano.2010.172}
  {\bibfield  {journal} {\bibinfo  {journal} {Nat Nano}\ }\textbf {\bibinfo
  {volume} {5}},\ \bibinfo {pages} {722} (\bibinfo {year} {2010})}\BibitemShut
  {NoStop}%
\bibitem [{\citenamefont {Mayorov}\ \emph {et~al.}(2011)\citenamefont
  {Mayorov}, \citenamefont {Gorbachev}, \citenamefont {Morozov}, \citenamefont
  {Britnell}, \citenamefont {Jalil}, \citenamefont {Ponomarenko}, \citenamefont
  {Blake}, \citenamefont {Novoselov}, \citenamefont {Watanabe}, \citenamefont
  {Taniguchi},\ and\ \citenamefont {Geim}}]{Mayorov_um_scale_balltr_GBN}%
  \BibitemOpen
  \bibfield  {author} {\bibinfo {author} {\bibfnamefont {A.~S.}\ \bibnamefont
  {Mayorov}}, \bibinfo {author} {\bibfnamefont {R.~V.}\ \bibnamefont
  {Gorbachev}}, \bibinfo {author} {\bibfnamefont {S.~V.}\ \bibnamefont
  {Morozov}}, \bibinfo {author} {\bibfnamefont {L.}~\bibnamefont {Britnell}},
  \bibinfo {author} {\bibfnamefont {R.}~\bibnamefont {Jalil}}, \bibinfo
  {author} {\bibfnamefont {L.~A.}\ \bibnamefont {Ponomarenko}}, \bibinfo
  {author} {\bibfnamefont {P.}~\bibnamefont {Blake}}, \bibinfo {author}
  {\bibfnamefont {K.~S.}\ \bibnamefont {Novoselov}}, \bibinfo {author}
  {\bibfnamefont {K.}~\bibnamefont {Watanabe}}, \bibinfo {author}
  {\bibfnamefont {T.}~\bibnamefont {Taniguchi}}, \ and\ \bibinfo {author}
  {\bibfnamefont {A.~K.}\ \bibnamefont {Geim}},\ }\href {\doibase
  10.1021/nl200758b} {\bibfield  {journal} {\bibinfo  {journal} {Nano Letters}\
  }\textbf {\bibinfo {volume} {11}},\ \bibinfo {pages} {2396} (\bibinfo {year}
  {2011})}\BibitemShut {NoStop}%
\bibitem [{\citenamefont {Zomer}\ \emph {et~al.}(2011)\citenamefont {Zomer},
  \citenamefont {Dash}, \citenamefont {Tombros},\ and\ \citenamefont {van
  Wees}}]{ZomerTransfer}%
  \BibitemOpen
  \bibfield  {author} {\bibinfo {author} {\bibfnamefont {P.~J.}\ \bibnamefont
  {Zomer}}, \bibinfo {author} {\bibfnamefont {S.~P.}\ \bibnamefont {Dash}},
  \bibinfo {author} {\bibfnamefont {N.}~\bibnamefont {Tombros}}, \ and\
  \bibinfo {author} {\bibfnamefont {B.~J.}\ \bibnamefont {van Wees}},\ }\href
  {\doibase 10.1063/1.3665405} {\bibfield  {journal}
  {\bibinfo  {journal} {Applied Physics Letters}\ }\textbf {\bibinfo {volume}
  {99}},\ \bibinfo {eid} {232104} (\bibinfo {year} {2011})}\BibitemShut
  {NoStop}%
\bibitem [{\citenamefont {Jalilian}\ \emph {et~al.}(2011)\citenamefont
  {Jalilian}, \citenamefont {Jauregui}, \citenamefont {Lopez}, \citenamefont
  {Tian}, \citenamefont {Roecker}, \citenamefont {Yazdanpanah}, \citenamefont
  {Cohn}, \citenamefont {Jovanovic},\ and\ \citenamefont
  {Chen}}]{JalilianAFMsgs}%
  \BibitemOpen
  \bibfield  {author} {\bibinfo {author} {\bibfnamefont {R.}~\bibnamefont
  {Jalilian}}, \bibinfo {author} {\bibfnamefont {L.~A.}\ \bibnamefont
  {Jauregui}}, \bibinfo {author} {\bibfnamefont {G.}~\bibnamefont {Lopez}},
  \bibinfo {author} {\bibfnamefont {J.}~\bibnamefont {Tian}}, \bibinfo {author}
  {\bibfnamefont {C.}~\bibnamefont {Roecker}}, \bibinfo {author} {\bibfnamefont
  {M.~M.}\ \bibnamefont {Yazdanpanah}}, \bibinfo {author} {\bibfnamefont
  {R.~W.}\ \bibnamefont {Cohn}}, \bibinfo {author} {\bibfnamefont
  {I.}~\bibnamefont {Jovanovic}}, \ and\ \bibinfo {author} {\bibfnamefont
  {Y.~P.}\ \bibnamefont {Chen}},\ }\href
  {http://stacks.iop.org/0957-4484/22/i=29/a=295705} {\bibfield  {journal}
  {\bibinfo  {journal} {Nanotechnology}\ }\textbf {\bibinfo {volume} {22}},\
  \bibinfo {pages} {295705} (\bibinfo {year} {2011})}\BibitemShut {NoStop}%
\bibitem [{\citenamefont {Goossens}\ \emph
  {et~al.}(2012{\natexlab{a}})\citenamefont {Goossens}, \citenamefont {Calado},
  \citenamefont {Barreiro}, \citenamefont {Watanabe}, \citenamefont
  {Taniguchi},\ and\ \citenamefont {Vandersypen}}]{GoossensAFMclean}%
  \BibitemOpen
  \bibfield  {author} {\bibinfo {author} {\bibfnamefont {A.~M.}\ \bibnamefont
  {Goossens}}, \bibinfo {author} {\bibfnamefont {V.~E.}\ \bibnamefont
  {Calado}}, \bibinfo {author} {\bibfnamefont {A.}~\bibnamefont {Barreiro}},
  \bibinfo {author} {\bibfnamefont {K.}~\bibnamefont {Watanabe}}, \bibinfo
  {author} {\bibfnamefont {T.}~\bibnamefont {Taniguchi}}, \ and\ \bibinfo
  {author} {\bibfnamefont {L.~M.~K.}\ \bibnamefont {Vandersypen}},\ }\href
  {\doibase 10.1063/1.3685504} {\bibfield  {journal}
  {\bibinfo  {journal} {Applied Physics Letters}\ }\textbf {\bibinfo {volume}
  {100}},\ \bibinfo {eid} {073110} (\bibinfo {year}
  {2012}{\natexlab{a}})}\BibitemShut {NoStop}%
\bibitem [{\citenamefont {Wang}\ \emph {et~al.}(2013)\citenamefont {Wang},
  \citenamefont {Meric}, \citenamefont {Huang}, \citenamefont {Gao},
  \citenamefont {Gao}, \citenamefont {Tran}, \citenamefont {Taniguchi},
  \citenamefont {Watanabe}, \citenamefont {Campos}, \citenamefont {Muller},
  \citenamefont {Guo}, \citenamefont {Kim}, \citenamefont {Hone}, \citenamefont
  {Shepard},\ and\ \citenamefont {Dean}}]{Wang1Dcontactto2Dmaterial}%
  \BibitemOpen
  \bibfield  {author} {\bibinfo {author} {\bibfnamefont {L.}~\bibnamefont
  {Wang}}, \bibinfo {author} {\bibfnamefont {I.}~\bibnamefont {Meric}},
  \bibinfo {author} {\bibfnamefont {P.~Y.}\ \bibnamefont {Huang}}, \bibinfo
  {author} {\bibfnamefont {Q.}~\bibnamefont {Gao}}, \bibinfo {author}
  {\bibfnamefont {Y.}~\bibnamefont {Gao}}, \bibinfo {author} {\bibfnamefont
  {H.}~\bibnamefont {Tran}}, \bibinfo {author} {\bibfnamefont {T.}~\bibnamefont
  {Taniguchi}}, \bibinfo {author} {\bibfnamefont {K.}~\bibnamefont {Watanabe}},
  \bibinfo {author} {\bibfnamefont {L.~M.}\ \bibnamefont {Campos}}, \bibinfo
  {author} {\bibfnamefont {D.~A.}\ \bibnamefont {Muller}}, \bibinfo {author}
  {\bibfnamefont {J.}~\bibnamefont {Guo}}, \bibinfo {author} {\bibfnamefont
  {P.}~\bibnamefont {Kim}}, \bibinfo {author} {\bibfnamefont {J.}~\bibnamefont
  {Hone}}, \bibinfo {author} {\bibfnamefont {K.~L.}\ \bibnamefont {Shepard}}, \
  and\ \bibinfo {author} {\bibfnamefont {C.~R.}\ \bibnamefont {Dean}},\ }\href
  {\doibase 10.1126/science.1244358} {\bibfield  {journal} {\bibinfo  {journal}
  {Science}\ }\textbf {\bibinfo {volume} {342}},\ \bibinfo {pages} {614}
  (\bibinfo {year} {2013})}\BibitemShut
  {NoStop}%
\bibitem [{\citenamefont {Zomer}\ \emph {et~al.}(2012)\citenamefont {Zomer},
  \citenamefont {Guimar\~aes}, \citenamefont {Tombros},\ and\ \citenamefont
  {van Wees}}]{ZomerSpinBN}%
  \BibitemOpen
  \bibfield  {author} {\bibinfo {author} {\bibfnamefont {P.~J.}\ \bibnamefont
  {Zomer}}, \bibinfo {author} {\bibfnamefont {M.~H.~D.}\ \bibnamefont
  {Guimar\~aes}}, \bibinfo {author} {\bibfnamefont {N.}~\bibnamefont
  {Tombros}}, \ and\ \bibinfo {author} {\bibfnamefont {B.~J.}\ \bibnamefont
  {van Wees}},\ }\href {\doibase 10.1103/PhysRevB.86.161416} {\bibfield
  {journal} {\bibinfo  {journal} {Phys. Rev. B}\ }\textbf {\bibinfo {volume}
  {86}},\ \bibinfo {pages} {161416} (\bibinfo {year} {2012})}\BibitemShut
  {NoStop}%
\bibitem [{\citenamefont {Jr.}\ \emph {et~al.}(2012)\citenamefont {Jr.},
  \citenamefont {Jing}, \citenamefont {Bao}, \citenamefont {Lee}, \citenamefont
  {Kratz}, \citenamefont {Aji}, \citenamefont {Bockrath}, \citenamefont {Lau},
  \citenamefont {Varma}, \citenamefont {Stillwell}, \citenamefont {Smirnov},
  \citenamefont {Zhang}, \citenamefont {Jung},\ and\ \citenamefont
  {MacDonald}}]{VelascoInsBLG}%
  \BibitemOpen
  \bibfield  {author} {\bibinfo {author} {\bibfnamefont {J.}\ \bibnamefont
  {Velasco~Jr.}}, \bibinfo {author} {\bibfnamefont {L.}~\bibnamefont {Jing}}, \bibinfo
  {author} {\bibfnamefont {W.}~\bibnamefont {Bao}}, \bibinfo {author}
  {\bibfnamefont {Y.}~\bibnamefont {Lee}}, \bibinfo {author} {\bibfnamefont
  {P.}~\bibnamefont {Kratz}}, \bibinfo {author} {\bibfnamefont
  {V.}~\bibnamefont {Aji}}, \bibinfo {author} {\bibfnamefont {M.}~\bibnamefont
  {Bockrath}}, \bibinfo {author} {\bibfnamefont {C.~N.}\ \bibnamefont {Lau}},
  \bibinfo {author} {\bibfnamefont {C.}~\bibnamefont {Varma}}, \bibinfo
  {author} {\bibfnamefont {R.}~\bibnamefont {Stillwell}}, \bibinfo {author}
  {\bibfnamefont {D.}~\bibnamefont {Smirnov}}, \bibinfo {author} {\bibfnamefont
  {F.}~\bibnamefont {Zhang}}, \bibinfo {author} {\bibfnamefont
  {J.}~\bibnamefont {Jung}}, \ and\ \bibinfo {author} {\bibfnamefont {A.~H.}\
  \bibnamefont {MacDonald}},\ }\href {http://dx.doi.org/10.1038/nnano.2011.251}
  {\bibfield  {journal} {\bibinfo  {journal} {Nat Nano}\ }\textbf {\bibinfo
  {volume} {7}},\ \bibinfo {pages} {156} (\bibinfo {year} {2012})}\BibitemShut
  {NoStop}%
\bibitem [{\citenamefont {Allen}, \citenamefont {Martin},\ and\ \citenamefont
  {Yacoby}(2012)}]{AllenQDinsuspBLG}%
  \BibitemOpen
  \bibfield  {author} {\bibinfo {author} {\bibfnamefont {M.~T.}\ \bibnamefont
  {Allen}}, \bibinfo {author} {\bibfnamefont {J.}~\bibnamefont {Martin}}, \
  and\ \bibinfo {author} {\bibfnamefont {A.}~\bibnamefont {Yacoby}},\ }\href
  {\doibase 10.1038/ncomms1945} {\bibfield  {journal} {\bibinfo  {journal} {Nat
  Commun}\ }\textbf {\bibinfo {volume} {3}},\ \bibinfo {pages} {934} (\bibinfo
  {year} {2012})}\BibitemShut {NoStop}%
\bibitem [{\citenamefont {van Elferen}\ \emph {et~al.}(2012)\citenamefont {van
  Elferen}, \citenamefont {Veligura}, \citenamefont {Kurganova}, \citenamefont
  {Zeitler}, \citenamefont {Maan}, \citenamefont {Tombros}, \citenamefont
  {Vera-Marun},\ and\ \citenamefont {van
  Wees}}]{vanElferen_supsBLG_QHferromagnetism}%
  \BibitemOpen
  \bibfield  {author} {\bibinfo {author} {\bibfnamefont {H.~J.}\ \bibnamefont
  {van Elferen}}, \bibinfo {author} {\bibfnamefont {A.}~\bibnamefont
  {Veligura}}, \bibinfo {author} {\bibfnamefont {E.~V.}\ \bibnamefont
  {Kurganova}}, \bibinfo {author} {\bibfnamefont {U.}~\bibnamefont {Zeitler}},
  \bibinfo {author} {\bibfnamefont {J.~C.}\ \bibnamefont {Maan}}, \bibinfo
  {author} {\bibfnamefont {N.}~\bibnamefont {Tombros}}, \bibinfo {author}
  {\bibfnamefont {I.~J.}\ \bibnamefont {Vera-Marun}}, \ and\ \bibinfo {author}
  {\bibfnamefont {B.~J.}\ \bibnamefont {van Wees}},\ }\href {\doibase
  10.1103/PhysRevB.85.115408} {\bibfield  {journal} {\bibinfo  {journal} {Phys.
  Rev. B}\ }\textbf {\bibinfo {volume} {85}},\ \bibinfo {pages} {115408}
  (\bibinfo {year} {2012})}\BibitemShut {NoStop}%
\bibitem [{\citenamefont {Veligura}\ \emph {et~al.}(2012)\citenamefont
  {Veligura}, \citenamefont {van Elferen}, \citenamefont {Tombros},
  \citenamefont {Maan}, \citenamefont {Zeitler},\ and\ \citenamefont {van
  Wees}}]{VeliguraTrGapinBLG0B}%
  \BibitemOpen
  \bibfield  {author} {\bibinfo {author} {\bibfnamefont {A.}~\bibnamefont
  {Veligura}}, \bibinfo {author} {\bibfnamefont {H.~J.}\ \bibnamefont {van
  Elferen}}, \bibinfo {author} {\bibfnamefont {N.}~\bibnamefont {Tombros}},
  \bibinfo {author} {\bibfnamefont {J.~C.}\ \bibnamefont {Maan}}, \bibinfo
  {author} {\bibfnamefont {U.}~\bibnamefont {Zeitler}}, \ and\ \bibinfo
  {author} {\bibfnamefont {B.~J.}\ \bibnamefont {van Wees}},\ }\href {\doibase
  10.1103/PhysRevB.85.155412} {\bibfield  {journal} {\bibinfo  {journal} {Phys.
  Rev. B}\ }\textbf {\bibinfo {volume} {85}},\ \bibinfo {pages} {155412}
  (\bibinfo {year} {2012})}\BibitemShut {NoStop}%
\bibitem [{\citenamefont {Goossens}\ \emph
  {et~al.}(2012{\natexlab{b}})\citenamefont {Goossens}, \citenamefont
  {Driessen}, \citenamefont {Baart}, \citenamefont {Watanabe}, \citenamefont
  {Taniguchi},\ and\ \citenamefont {Vandersypen}}]{GoossensQdot}%
  \BibitemOpen
  \bibfield  {author} {\bibinfo {author} {\bibfnamefont {A.~S.~M.}\
  \bibnamefont {Goossens}}, \bibinfo {author} {\bibfnamefont {S.~C.~M.}\
  \bibnamefont {Driessen}}, \bibinfo {author} {\bibfnamefont {T.~A.}\
  \bibnamefont {Baart}}, \bibinfo {author} {\bibfnamefont {K.}~\bibnamefont
  {Watanabe}}, \bibinfo {author} {\bibfnamefont {T.}~\bibnamefont {Taniguchi}},
  \ and\ \bibinfo {author} {\bibfnamefont {L.~M.~K.}\ \bibnamefont
  {Vandersypen}},\ }\href {\doibase 10.1021/nl301986q} {\bibfield  {journal}
  {\bibinfo  {journal} {Nano Letters}\ }\textbf {\bibinfo {volume} {12}},\
  \bibinfo {pages} {4656} (\bibinfo {year} {2012}{\natexlab{b}})}\BibitemShut {NoStop}%
\bibitem [{\citenamefont {Haigh}\ \emph {et~al.}(2012)\citenamefont {Haigh},
  \citenamefont {Gholinia}, \citenamefont {Jalil}, \citenamefont {Romani},
  \citenamefont {Britnell}, \citenamefont {Elias}, \citenamefont {Novoselov},
  \citenamefont {Ponomarenko}, \citenamefont {Geim},\ and\ \citenamefont
  {Gorbachev}}]{Haigh_NatMat_bubblesGhBN}%
  \BibitemOpen
  \bibfield  {author} {\bibinfo {author} {\bibfnamefont {S.~J.}\ \bibnamefont
  {Haigh}}, \bibinfo {author} {\bibfnamefont {A.}~\bibnamefont {Gholinia}},
  \bibinfo {author} {\bibfnamefont {R.}~\bibnamefont {Jalil}}, \bibinfo
  {author} {\bibfnamefont {S.}~\bibnamefont {Romani}}, \bibinfo {author}
  {\bibfnamefont {L.}~\bibnamefont {Britnell}}, \bibinfo {author}
  {\bibfnamefont {D.~C.}\ \bibnamefont {Elias}}, \bibinfo {author}
  {\bibfnamefont {K.~S.}\ \bibnamefont {Novoselov}}, \bibinfo {author}
  {\bibfnamefont {L.~A.}\ \bibnamefont {Ponomarenko}}, \bibinfo {author}
  {\bibfnamefont {A.~K.}\ \bibnamefont {Geim}}, \ and\ \bibinfo {author}
  {\bibfnamefont {R.}~\bibnamefont {Gorbachev}},\ }\href {\doibase
  10.1038/nmat3386} {\bibfield  {journal} {\bibinfo  {journal} {Nat Mater}\
  }\textbf {\bibinfo {volume} {11}},\ \bibinfo {pages} {764 } (\bibinfo {year}
  {2012})}\BibitemShut {NoStop}%
\bibitem [{\citenamefont {Xue}\ \emph {et~al.}(2011)\citenamefont {Xue},
  \citenamefont {Sanchez-Yamagishi}, \citenamefont {Bulmash}, \citenamefont
  {Jacquod}, \citenamefont {Deshpande}, \citenamefont {Watanabe}, \citenamefont
  {Taniguchi}, \citenamefont {Jarillo-Herrero},\ and\ \citenamefont
  {LeRoy}}]{XueSTM_GBN}%
  \BibitemOpen
  \bibfield  {author} {\bibinfo {author} {\bibfnamefont {J.}~\bibnamefont
  {Xue}}, \bibinfo {author} {\bibfnamefont {J.}~\bibnamefont
  {Sanchez-Yamagishi}}, \bibinfo {author} {\bibfnamefont {D.}~\bibnamefont
  {Bulmash}}, \bibinfo {author} {\bibfnamefont {P.}~\bibnamefont {Jacquod}},
  \bibinfo {author} {\bibfnamefont {A.}~\bibnamefont {Deshpande}}, \bibinfo
  {author} {\bibfnamefont {K.}~\bibnamefont {Watanabe}}, \bibinfo {author}
  {\bibfnamefont {T.}~\bibnamefont {Taniguchi}}, \bibinfo {author}
  {\bibfnamefont {P.}~\bibnamefont {Jarillo-Herrero}}, \ and\ \bibinfo {author}
  {\bibfnamefont {B.~J.}\ \bibnamefont {LeRoy}},\ }\href
  {http://dx.doi.org/10.1038/nmat2968} {\bibfield  {journal} {\bibinfo
  {journal} {Nat Mater}\ }\textbf {\bibinfo {volume} {10}},\ \bibinfo {pages}
  {282} (\bibinfo {year} {2011})}\BibitemShut {NoStop}%
\bibitem [{\citenamefont {Decker}\ \emph {et~al.}(2011)\citenamefont {Decker},
  \citenamefont {Wang}, \citenamefont {Brar}, \citenamefont {Regan},
  \citenamefont {Tsai}, \citenamefont {Wu}, \citenamefont {Gannett},
  \citenamefont {Zettl},\ and\ \citenamefont {Crommie}}]{Decker_STM_GBN}%
  \BibitemOpen
  \bibfield  {author} {\bibinfo {author} {\bibfnamefont {R.}~\bibnamefont
  {Decker}}, \bibinfo {author} {\bibfnamefont {Y.}~\bibnamefont {Wang}},
  \bibinfo {author} {\bibfnamefont {V.~W.}\ \bibnamefont {Brar}}, \bibinfo
  {author} {\bibfnamefont {W.}~\bibnamefont {Regan}}, \bibinfo {author}
  {\bibfnamefont {H.-Z.}\ \bibnamefont {Tsai}}, \bibinfo {author}
  {\bibfnamefont {Q.}~\bibnamefont {Wu}}, \bibinfo {author} {\bibfnamefont
  {W.}~\bibnamefont {Gannett}}, \bibinfo {author} {\bibfnamefont
  {A.}~\bibnamefont {Zettl}}, \ and\ \bibinfo {author} {\bibfnamefont {M.~F.}\
  \bibnamefont {Crommie}},\ }\href {\doibase 10.1021/nl2005115} {\bibfield
  {journal} {\bibinfo  {journal} {Nano Letters}\ }\textbf {\bibinfo {volume}
  {11}},\ \bibinfo {pages} {2291} (\bibinfo {year} {2011})}\BibitemShut {NoStop}%
\bibitem [{\citenamefont {Yankowitz}\ \emph {et~al.}(2012)\citenamefont
  {Yankowitz}, \citenamefont {Xue}, \citenamefont {Cormode}, \citenamefont
  {Sanchez-Yamagishi}, \citenamefont {Watanabe}, \citenamefont {Taniguchi},
  \citenamefont {Jarillo-Herrero}, \citenamefont {Jacquod},\ and\ \citenamefont
  {LeRoy}}]{YankowitzGBNsuperlatticeSTM}%
  \BibitemOpen
  \bibfield  {author} {\bibinfo {author} {\bibfnamefont {M.}~\bibnamefont
  {Yankowitz}}, \bibinfo {author} {\bibfnamefont {J.}~\bibnamefont {Xue}},
  \bibinfo {author} {\bibfnamefont {D.}~\bibnamefont {Cormode}}, \bibinfo
  {author} {\bibfnamefont {J.~D.}\ \bibnamefont {Sanchez-Yamagishi}}, \bibinfo
  {author} {\bibfnamefont {K.}~\bibnamefont {Watanabe}}, \bibinfo {author}
  {\bibfnamefont {T.}~\bibnamefont {Taniguchi}}, \bibinfo {author}
  {\bibfnamefont {P.}~\bibnamefont {Jarillo-Herrero}}, \bibinfo {author}
  {\bibfnamefont {P.}~\bibnamefont {Jacquod}}, \ and\ \bibinfo {author}
  {\bibfnamefont {B.~J.}\ \bibnamefont {LeRoy}},\ }\href {\doibase
  10.1038/nphys2272} {\bibfield  {journal} {\bibinfo  {journal} {Nature
  Physics}\ }\textbf {\bibinfo {volume} {8}},\ \bibinfo {pages} {382} (\bibinfo
  {year} {2012})}\BibitemShut {NoStop}%
\bibitem [{\citenamefont {Ponomarenko}\ \emph {et~al.}(2013)\citenamefont
  {Ponomarenko}, \citenamefont {Gorbachev}, \citenamefont {Yu}, \citenamefont
  {Elias}, \citenamefont {Jalil}, \citenamefont {Patel}, \citenamefont
  {Mishchenko}, \citenamefont {Mayorov}, \citenamefont {Woods}, \citenamefont
  {Wallbank}, \citenamefont {Mucha-Kruczynski}, \citenamefont {Piot},
  \citenamefont {Potemski}, \citenamefont {Grigorieva}, \citenamefont
  {Novoselov}, \citenamefont {Guinea}, \citenamefont {Fal’ko},\ and\
  \citenamefont {Geim}}]{PonomarenkoHBF}%
  \BibitemOpen
  \bibfield  {author} {\bibinfo {author} {\bibfnamefont {L.~A.}\ \bibnamefont
  {Ponomarenko}}, \bibinfo {author} {\bibfnamefont {R.~V.}\ \bibnamefont
  {Gorbachev}}, \bibinfo {author} {\bibfnamefont {G.~L.}\ \bibnamefont {Yu}},
  \bibinfo {author} {\bibfnamefont {D.~C.}\ \bibnamefont {Elias}}, \bibinfo
  {author} {\bibfnamefont {R.}~\bibnamefont {Jalil}}, \bibinfo {author}
  {\bibfnamefont {A.~A.}\ \bibnamefont {Patel}}, \bibinfo {author}
  {\bibfnamefont {A.}~\bibnamefont {Mishchenko}}, \bibinfo {author}
  {\bibfnamefont {A.~S.}\ \bibnamefont {Mayorov}}, \bibinfo {author}
  {\bibfnamefont {C.~R.}\ \bibnamefont {Woods}}, \bibinfo {author}
  {\bibfnamefont {J.~R.}\ \bibnamefont {Wallbank}}, \bibinfo {author}
  {\bibfnamefont {M.}~\bibnamefont {Mucha-Kruczynski}}, \bibinfo {author}
  {\bibfnamefont {B.~A.}\ \bibnamefont {Piot}}, \bibinfo {author}
  {\bibfnamefont {M.}~\bibnamefont {Potemski}}, \bibinfo {author}
  {\bibfnamefont {I.~V.}\ \bibnamefont {Grigorieva}}, \bibinfo {author}
  {\bibfnamefont {K.~S.}\ \bibnamefont {Novoselov}}, \bibinfo {author}
  {\bibfnamefont {F.}~\bibnamefont {Guinea}}, \bibinfo {author} {\bibfnamefont
  {V.~I.}\ \bibnamefont {Fal’ko}}, \ and\ \bibinfo {author} {\bibfnamefont
  {A.~K.}\ \bibnamefont {Geim}},\ }\href {\doibase 10.1038/nature12187}
  {\bibfield  {journal} {\bibinfo  {journal} {Nature}\ }\textbf {\bibinfo
  {volume} {497}},\ \bibinfo {pages} {594} (\bibinfo {year}
  {2013})}\BibitemShut {NoStop}%
\bibitem [{\citenamefont {Hunt}\ \emph {et~al.}(2013)\citenamefont {Hunt},
  \citenamefont {Sanchez-Yamagishi}, \citenamefont {Young}, \citenamefont
  {Yankowitz}, \citenamefont {LeRoy}, \citenamefont {Watanabe}, \citenamefont
  {Taniguchi}, \citenamefont {Moon}, \citenamefont {Koshino}, \citenamefont
  {Jarillo-Herrero},\ and\ \citenamefont {Ashoori}}]{HuntHBF}%
  \BibitemOpen
  \bibfield  {author} {\bibinfo {author} {\bibfnamefont {B.}~\bibnamefont
  {Hunt}}, \bibinfo {author} {\bibfnamefont {J.~D.}\ \bibnamefont
  {Sanchez-Yamagishi}}, \bibinfo {author} {\bibfnamefont {A.~F.}\ \bibnamefont
  {Young}}, \bibinfo {author} {\bibfnamefont {M.}~\bibnamefont {Yankowitz}},
  \bibinfo {author} {\bibfnamefont {B.~J.}\ \bibnamefont {LeRoy}}, \bibinfo
  {author} {\bibfnamefont {K.}~\bibnamefont {Watanabe}}, \bibinfo {author}
  {\bibfnamefont {T.}~\bibnamefont {Taniguchi}}, \bibinfo {author}
  {\bibfnamefont {P.}~\bibnamefont {Moon}}, \bibinfo {author} {\bibfnamefont
  {M.}~\bibnamefont {Koshino}}, \bibinfo {author} {\bibfnamefont
  {P.}~\bibnamefont {Jarillo-Herrero}}, \ and\ \bibinfo {author} {\bibfnamefont
  {R.~C.}\ \bibnamefont {Ashoori}},\ }\href {\doibase 10.1126/science.1237240}
  {\bibfield  {journal} {\bibinfo  {journal} {Science}\ }\textbf {\bibinfo
  {volume} {340}},\ \bibinfo {pages} {1427} (\bibinfo {year} {2013})}\BibitemShut
  {NoStop}%
\bibitem [{\citenamefont {Dean}\ \emph {et~al.}(2013)\citenamefont {Dean},
  \citenamefont {Wang}, \citenamefont {Maher}, \citenamefont {Forsythe},
  \citenamefont {Ghahari}, \citenamefont {Gao}, \citenamefont {Katoch},
  \citenamefont {Ishigami}, \citenamefont {Moon}, \citenamefont {Koshino},
  \citenamefont {Taniguchi}, \citenamefont {Watanabe}, \citenamefont {Shepard},
  \citenamefont {Hone},\ and\ \citenamefont {Kim}}]{DeanHBF}%
  \BibitemOpen
  \bibfield  {author} {\bibinfo {author} {\bibfnamefont {C.~R.}\ \bibnamefont
  {Dean}}, \bibinfo {author} {\bibfnamefont {L.}~\bibnamefont {Wang}}, \bibinfo
  {author} {\bibfnamefont {P.}~\bibnamefont {Maher}}, \bibinfo {author}
  {\bibfnamefont {C.}~\bibnamefont {Forsythe}}, \bibinfo {author}
  {\bibfnamefont {F.}~\bibnamefont {Ghahari}}, \bibinfo {author} {\bibfnamefont
  {Y.}~\bibnamefont {Gao}}, \bibinfo {author} {\bibfnamefont {J.}~\bibnamefont
  {Katoch}}, \bibinfo {author} {\bibfnamefont {M.}~\bibnamefont {Ishigami}},
  \bibinfo {author} {\bibfnamefont {P.}~\bibnamefont {Moon}}, \bibinfo {author}
  {\bibfnamefont {M.}~\bibnamefont {Koshino}}, \bibinfo {author} {\bibfnamefont
  {T.}~\bibnamefont {Taniguchi}}, \bibinfo {author} {\bibfnamefont
  {K.}~\bibnamefont {Watanabe}}, \bibinfo {author} {\bibfnamefont {K.~L.}\
  \bibnamefont {Shepard}}, \bibinfo {author} {\bibfnamefont {J.}~\bibnamefont
  {Hone}}, \ and\ \bibinfo {author} {\bibfnamefont {P.}~\bibnamefont {Kim}},\
  }\href {\doibase 10.1038/nature12186} {\bibfield  {journal} {\bibinfo
  {journal} {Nature}\ }\textbf {\bibinfo {volume} {497}},\ \bibinfo {pages}
  {598} (\bibinfo {year} {2013})}\BibitemShut {NoStop}%
\bibitem [{\citenamefont {Pease}(1952)}]{PeaseXrayBN}%
  \BibitemOpen
  \bibfield  {author} {\bibinfo {author} {\bibfnamefont {R.}~\bibnamefont
  {Pease}},\ }\href {\doibase 10.1103/PhysRevB.80.235402} {\bibfield  {journal}
  {\bibinfo  {journal} {Acta Cryst.}\ }\textbf {\bibinfo {volume} {5}},\
  \bibinfo {pages} {356} (\bibinfo {year} {1952})}\BibitemShut {NoStop}%
\bibitem [{\citenamefont {Bolotin}\ \emph {et~al.}(2008)\citenamefont
  {Bolotin}, \citenamefont {Sikes}, \citenamefont {Jiang}, \citenamefont
  {Klima}, \citenamefont {Fudenberg}, \citenamefont {Hone}, \citenamefont
  {Kim},\ and\ \citenamefont
  {Stormer}}]{SSC146_Bolotin2008_UltrahighElectronMobilityinSuspendedGraphene}%
  \BibitemOpen
  \bibfield  {author} {\bibinfo {author} {\bibfnamefont {K.}~\bibnamefont
  {Bolotin}}, \bibinfo {author} {\bibfnamefont {K.}~\bibnamefont {Sikes}},
  \bibinfo {author} {\bibfnamefont {Z.}~\bibnamefont {Jiang}}, \bibinfo
  {author} {\bibfnamefont {M.}~\bibnamefont {Klima}}, \bibinfo {author}
  {\bibfnamefont {G.}~\bibnamefont {Fudenberg}}, \bibinfo {author}
  {\bibfnamefont {J.}~\bibnamefont {Hone}}, \bibinfo {author} {\bibfnamefont
  {P.}~\bibnamefont {Kim}}, \ and\ \bibinfo {author} {\bibfnamefont
  {H.}~\bibnamefont {Stormer}},\ }\href {\doibase DOI:
  10.1016/j.ssc.2008.02.024} {\bibfield  {journal} {\bibinfo  {journal} {Solid
  State Communications}\ }\textbf {\bibinfo {volume} {146}},\ \bibinfo {pages}
  {351 } (\bibinfo {year} {2008})}\BibitemShut {NoStop}%
\bibitem [{\citenamefont {Elias}\ \emph {et~al.}(2011)\citenamefont {Elias},
  \citenamefont {Gorbachev}, \citenamefont {Mayorov}, \citenamefont {Morozov},
  \citenamefont {Zhukov}, \citenamefont {Blake}, \citenamefont {Ponomarenko},
  \citenamefont {Grigorieva}, \citenamefont {Novoselov}, \citenamefont
  {Guinea},\ and\ \citenamefont {Geim}}]{Elias_DiracConeReshaped_suspSLG}%
  \BibitemOpen
  \bibfield  {author} {\bibinfo {author} {\bibfnamefont {D.~C.}\ \bibnamefont
  {Elias}}, \bibinfo {author} {\bibfnamefont {R.~V.}\ \bibnamefont
  {Gorbachev}}, \bibinfo {author} {\bibfnamefont {A.~S.}\ \bibnamefont
  {Mayorov}}, \bibinfo {author} {\bibfnamefont {S.~V.}\ \bibnamefont
  {Morozov}}, \bibinfo {author} {\bibfnamefont {A.~A.}\ \bibnamefont {Zhukov}},
  \bibinfo {author} {\bibfnamefont {P.}~\bibnamefont {Blake}}, \bibinfo
  {author} {\bibfnamefont {L.~A.}\ \bibnamefont {Ponomarenko}}, \bibinfo
  {author} {\bibfnamefont {I.~V.}\ \bibnamefont {Grigorieva}}, \bibinfo
  {author} {\bibfnamefont {K.~S.}\ \bibnamefont {Novoselov}}, \bibinfo {author}
  {\bibfnamefont {F.}~\bibnamefont {Guinea}}, \ and\ \bibinfo {author}
  {\bibfnamefont {A.~K.}\ \bibnamefont {Geim}},\ }\href
  {http://dx.doi.org/10.1038/nphys2049} {\bibfield  {journal} {\bibinfo
  {journal} {Nat Phys}\ }\textbf {\bibinfo {volume} {7}},\ \bibinfo {pages}
  {701} (\bibinfo {year} {2011})}\BibitemShut {NoStop}%
\end{thebibliography}

%

\end{document}